\documentclass[article]{rmaa}

\usepackage{paralist}

\usepackage{psfrag,color,psfig}

\SetYear{}
\SetConfTitle{}

\title{The Kinematics of Arp 295 in $\rm H\alpha$ Emission: an Interacting Galaxy with Highly Asymmetric Rotation}

\author{
  Nathan Roche,\altaffilmark{1}}

\altaffiltext{1}{OAN, Instituto de Astronom\'\i{}a, UNAM, Ensenada, M\'exico. Now at SAAO, South Africa.}

\shortauthor{Roche N.D.}
\shorttitle{$\rm H\alpha$ Kinematics of Arp 295}

\fulladdresses{
\item Nathan Roche: Observatorio Astronomico Nacional,
 Instituto de Astronom\'ia, Universidad Nacional
  Aut\'onoma de M\'exico, Ensenada, Apartado Postal 3--72, 22860
  Ensenada, Baja California, M\'exico.
\item now at: South African Astronomical Observatory, PO Box 9,
Observatory 7935, Cape Town, South Africa (\email {roche@ukzn.ac.za}).
}

\listofauthors{N.D. Roche}
\indexauthor{Roche, N.D.}

\abstract{We investigate Arp 295, a pair of interacting spirals
 at $z=0.023$, using optical spectroscopy, $\rm H\alpha$ imaging, and $\rm H\alpha$ velocity mapping with the Manchester Echelle Spectrograph.
  Scalelengths are 
$r_{exp}=5.24$ for Arp 295a and 2.52 kpc for 295b. A much smaller Im galaxy, Arp 295c, is associated with the larger spiral. 
Arp 295b is asymmetric with the disk more extended eastwards but with the brightest star-formation regions on its western side. 
The spectrum of Arp 295b shows strong emission lines with [OII]3727 equivalent width $36\rm \AA$.  The total $\rm H\alpha$ luminosities of Arp 295b and Arp 295c are $4.69\times 10^{41}$ and
$6.76\times 10^{40}$ ergs $\rm s^{-1}$, corresponding to star-formation rates 3.7 and 0.53 $\rm M_{\odot} yr^{-1}$.  
For Arp 295b, we measure the maximum disk rotation velocity as $252.6\pm 9.9$ km $\rm s^{-1}$ and find the 
rotation curve is very asymmetric.
  The east (approaching) side has a higher radial velocity than the west with the maximum difference (at $r=5$ arcsec) 88 km $\rm s^{-1}$ or a factor 1.675. 

}

\addkeyword{Galaxies: Interacting}

\begin{document}
\maketitle

\section{Introduction}
\label{sec:intro}
 Arp 295, listed in the
 catalog of Arp (1966) as a `double galaxy with long filaments', consists primarily of two interacting spiral galaxies:
 the south-western, Arp 295a, is a
  nearly edge-on (inclined $85^{\circ}$) Sc-type, the north-eastern, Arp 295b
  is an inclined ($55^{\circ}$) Sb/peculiar/HII. There are visible tidal features -- a luminous bridge connecting the two galaxies, a
  a long luminous tail extending further SW from 295a, and a broad plume extending eastwards from Arp 295b.
  
Arp 295a is at RA $23^h 41^m 47^s.3$, Dec -03:40:02 and Arp 295b lies 4.59
arcmin away at RA $23^h 42^m 00^s.8$, Dec -03:36:55. The
redshifts (recession velocities) are given by de Vaucouleurs et al.
(1991) as  $z=0.022846$ ($6849\pm 20$ km $\rm s^{-1}$) for 295a
 and $z=0.023239$ ($6967\pm 20$  km $\rm s^{-1}$) for 295b, greater by $118\pm 28$ km $\rm s^{-1}$. For $H_0=70$ km $\rm s^{-1}$ (assumed throughout this paper),
the proper distance is 99.1 Mpc, the distance modulus 35.031 mag, and 1 arcsec will subtend
469.7 pc.

Stockton (1974) obtained long-slit spectra for the two
galaxies, measured their radial velocity difference as
$100\pm 20$ km $\rm s^{-1}$, found Arp 295b is rotating in the sense that it is receding on its western side, and that Arp 295a is receding on
its northeastern. Thus with respect to their mutual orbit, Arp 295b rotates retrograde and Arp 295a prograde. Stockton
 found maximum rotation velocities (inclination-corrected) of 300 km $\rm s^{-1}$ for Arp 295a
 (with a turnover at $r=40$ arcsec) and
 245 km $\rm s^{-1}$ for Arp 295b (turnover at $r=17$ arcsec), estimating from this that the 295a component is $\sim 3.5$ times 
 more massive. However, their $B$-band magnitudes are almost the same, $B=14.50/14.60$ for Arp 295a/b (de Vaucouleurs et al. 1991).

Arp 295 forms the
first (earliest stage) in the sequence of mergers studied optically and in HI by Hibbard and van Gorkom (1996). They give far infra-red luminosities   $4.69\times 10^{43}$ ergs $\rm s^{-1}$ 295a and $3.83\times 10^{44}$
 ergs $\rm s^{-1}$ for 295b, which from the relation of Kennicutt (1998), correspond to star-formation rates (SFRs) of
2.1 and 17.2$\rm ~M_{\odot}yr^{-1}$. They find that the more active Arp295b is also more gas-rich than the larger galaxy, with $\rm M_{HI}\simeq 1.69\times 10^{10}~~M_{\odot}$ compared to
 $\rm 4.9\times 10^{9}~M_{\odot}$.

 Dopita et al. (2002) measure a $\rm H\alpha$ flux of
 $3.13\times 10^{-13}$ ergs $\rm cm^{-2} s^{-1}$ for Arp 295b, which corresponds to
  $L_{\rm H\alpha}=3.85\times 10^{41}$ ergs
 $\rm s^{-1}$, and from the relation of Kennicutt (1998),
  ${\rm SFR}=7.9\times 10^{-42}L_{\rm H\alpha}$, a SFR of
 3.04$\rm ~M_{\odot}yr^{-1}$. Most of the $\rm H\alpha$ emission is from a number of `knots' within the galaxy disk, with fainter ring-shaped emission to the north and south.
From the far infra-red luminosity they derive a much higher SFR of 22.8$\rm ~M_{\odot}yr^{-1}$
  and account for  the difference as dust extinction.

  Neff et al.(2005), using GALEX data, detected UV emission from two small clumps within the gas-rich
  eastern plume of Arp 295b, about 90 arcsec ENE of the nucleus. These have very blue UV-optical colours, suggesting stellar ages $\sim 10^7$, and it was 
  speculated that
   they could be formative tidal dwarf galaxies, each containing $\sim 10^6~\rm M_{\odot}$ of stars but $\sim 10^8~\rm M_{\odot}$ of gas.

Keel (1993) studied the kinematics of a sample of interacting pairs of spirals and found an association between
kinematic disturbance, in the form of unusually low rotation velocities, and triggered star-formation.  
In this paper, we study kinematics and star-formation in the Arp 295 
system by means of optical spectroscopy, deep imaging in $\rm H\alpha$, and 
high-resolution spectroscopy of the $\rm H\alpha$ line with the Manchester Echelle Spectrograph (MES).

 In Section 2 we describe our observations and
data reduction. In Section 3 we present and interpret the optical spectroscopy, and  in Section 4
the direct imaging in $\rm H\alpha$. In Section 5 we use the
$\rm H\alpha$ echelle spectroscopy to derive a velocity map and rotation curve.
 In Section 6 we summarize and discuss the effects of the galaxy interaction as revealed in our data.

\section{Observations and Data}
\label{sec:obs}
\subsection{Observations at OAN-SPM}
All observations were carried OUT at the Observatorio
Astronomico Nacional,
 in the Sierra San Pedro Martir,
Baja California, M\'exico. The optical spectroscopy was performed on the night of 14 August 2005,
using the `Bolitas' spectrograph on the 0.84m telescope. This small Boller and Chivens spectrograph
was fitted with a Thompson 2k CCD (operated with $2\times2$ binning to give a
 $1024\times 1024$ pixel image), an RGL grating set at an angle of approximately $6^{\circ}$
to cover $3600\leq \lambda \leq 5920\rm \AA$, and a
slit of width $160\rm \mu m$ aligned E-W.
 To obtain a relative
flux calibration we observed (through the same slit) the
 spectrophotometric standard star
 58 Aquilae (Hamuy et al. 1994). For wavelength calibration, spectra were taken of a
HeAr arc lamp. 

Due to weather conditions and other difficulties we obtained usable
spectra for only the more active galaxy, Arp 295b, consisting of  
 $6\times 1200$s exposures.

Our observations in $\rm H\alpha$ (restframe $\lambda=6562.801\rm \AA$ in air) were performed on the nights of 15--18 August 2005, using the Manchester Echelle
Spectrograph (MES) on the 2.1m telescope. The MES (Meaburn 2003) is designed to obtain
spatially-resolved profiles of individual emission lines, from faint extended sources such as nebulae and galaxies. It
achieves high spectral/velocity resolution ($>10^5$) and throughput 
 by operating in a high spectral order with a narrow-band filter but no cross-disperser.

MES is fitted with a SITe3 CCD, which was operated with $2\times2$ binning to give a $512\times 512$ pixel image
of pixelsize 0.62 arcsec. The gain is $1.27e^{-1}$/ADU and we measure the readout noise as 12.35 ADU. 
At the redshift of Arp 295b, $\rm H\alpha$ is redshifted to $6715\rm \AA$, where it falls near the
centre (where sensitivity is greatest) of the echelle's 89th order. To isolate the redshifted $\rm H\alpha$ we observed
 through the E6690 filter, with width $91\rm \AA$.

 Three types of observation were taken with MES. Firstly, the instrument was used in direct-imaging mode -- a
  mirror is inserted to bypass the slit and grating and image directly on to the CCD. The effective field of view is
   approximately $5.3\times 4.1$ arcmin, so separate pointings are required for Arp 295a and 295b.
    In  E6690, we obtained $5\times 1200$s exposures of Arp 295b and one 1200s exposure of Arp 295a. For flux
   calibration, direct imaging was taken of the faint
   spectrophotometric standard star BD+33 2642 (Oke 1990). 

   Secondly, observation were taken though the slit and grating, to obtain a
   high-resolution profile of the $\rm H\alpha$ emission line at all points on the target
   galaxy along the slit. The slit, width $\rm 150\mu m$, was aligned E-W, and a total of $9\times 1200$s exposures were taken on Arp 295b. The slit was moved slightly in the N-S direction between each exposure, so to cover 9
    positions from the north to the south edge of the galaxy disk. Several further 1200s exposures were taken with the slit on Arp 295a or on other objects. 
        Immediately before every exposure in spectroscopic mode, a 60 second `mirror-slit' image was taken, which
	shows the slit position superimposed on a direct image of the galaxy, and a Th-Ar arc lamp exposure was taken for wavelength
	 calibration.

	Thirdly, exposures were taken through the grating with an open aperture instead of a slit, one of 1200s on each of Arp 295a and 295b.   In this mode the $\rm H\alpha$ line will be
	`smeared' by the extent of the galaxy in the dispersion direction. The motivation for this is that the integrated, background-subtracted flux in the slitless $\rm H\alpha$ line profile
	will give the total $\rm H\alpha$ flux of
	 the galaxy without contamination from the red continuum. For flux calibration a slitless spectrum was taken of (again) 58 Aquilae.
\subsection{Data Reduction: `Bolitas'}
\label{sec:reduc1}
The `Bolitas' spectroscopic data were debiased and  flat-fielded.
Twilight sky flats were taken through the same slit/grating
configuration as used for the observations.
In the flat-fielding of spectroscopic data the aim is to
 remove the pixel-to-pixel sensitivity
 variations but not the variation along the
wavelength direction. To do this, each pixel
 of the sky-flat was divided by the average value in its column
 (i.e. of all pixels at the
 same wavelength, averaged along the slit length), to produce a normalized
flat-field with all wavelength dependence removed.

Calibration spectra of 58 Aquilae were
spatially registered
and combined, the 1D spectrum extracted (using IRAF
`apall'), wavelength-calibrated, and, with the tabulated spectral energy distribution for
this star (Hamuy et al. 1994), used
 to derive a
sensitivity (relative flux calibration) function.
Our $6\times 1200$s exposures of Arp 295b were spatially
registered and combined (with `sigclip' cosmic-ray rejection), then the
1D spectrum was extracted,
wavelength calibrated, and flux calibrated in
relative $F_{\lambda}$.
\subsection{Data Reduction: MES}
\label{sec:reduc2}
All the data were debiased. The twilight flats in direct-imaging mode were combined and normalized to
 give a master sky-flat for each observing night. All direct imaging and spectroscopic observations were then
 divided by the flat-field for their night of observation.

The flux of our calibration standard BD+33:2642 was measured using IRAF aperture photometry. From the tabulated spectral energy distribution of Oke (1990), the AB magnitude in
 the E6690 filter wavelength range is 11.080. We derive a zero-point for our E6690 direct-imaging of
 $21.209-0.089X$, where $X$ is the airmass. A further correction is required for Galactic extinction at these co-ordinates - according to the Caltech NED, 0.102 mag at $0.65\rm \mu m$ and 0.165
mag in the $B$ band. We use this value but note that it could be an overestimate as Burstein and Heiles (1982) estimate extinction almost a factor of 2 smaller. This gives a zeropoint 20.981 for our combined observations of Arp 295b, and 20.998 for Arp 295a.

 Our $5\times 1200$s  direct-imaging exposures of Arp 295b were spatially registered, sky subtracted, and combined, with
 `sigclip' cosmic-ray removal. This was not fully effective in removing all cosmic rays, so an additional cleaning was performed with  IRAF `crmedian'. Our single 1200s
exposure of Arp 295a was simply sky-subtracted and cleaned with `crmedian'. 

The spectroscopic mode MES images were also cleaned with `crmedian'. and each was wavelength calibrated with the associated ThAr arc (we identified the lines with the NOAO
 spectral atlas). We find that our observations cover the range 6693.9 to $6745.7\rm \AA$
 with a dispersion $0.1014\rm \AA~pixel^{-1}$.
From our slitless spectrum of 58 Aquilae and the tabulated spectral energy distribution of Oke (1990), we derived an
absolute flux calibration, which indicated the sensitivity of MES in our configuration to
peak at $6725.5\rm \AA$ and vary by $\leq 0.5$ mag over the observed $\lambda$ range.

\section{Optical Spectroscopy}
\label{sec:spec}
Figure 1 shows the optical spectrum of Arp 295b, from 7200s of observation with Bolitas. The galaxy is bright in the UV 
and has strong emission lines, especially [OII]3727, indicating high star-formation activity. The spectrum is plotted
with  only a relative flux calibration, as the slit will capture only a fraction of the flux
from the galaxy.
 However, if it is assumed that the region covered by the slit is representative
  of the whole galaxy, we can estimate 
 an approximate absolute calibration by integrating the
 observed spectrum over a standard
 $B$-band response function and equating to the total
 magnitude $B\simeq 14.50$, which is $F_{\lambda}\simeq 9.53\times 10^{-15}$
ergs $\rm cm^{-2}s^{-1}$. In this way we estimate that one flux unit
 on Figure 1 corresponds to a flux from the whole galaxy of
 $3.10\times 10^{-15}$ ergs $\rm cm^{-2} s^{-1} \AA^{-1}$, and a 
 luminosity $3.82\times 10^{39}$ ergs $\rm s^{-1} \AA^{-1}$, 

From our spectrum we calculate the $B$-band k-correction as 0.013
magnitudes. Using this, the absolute magnitude is $M_B=-20.44$. 

Examining the spectrum we find 5 significant emission lines and 4 absorption lines. From the wavelengths of the emission lines we derive a redshift
$z=0.02322\pm 0.00005$, or recession velocity $6961\pm15$ km $\rm s^{-1}$, in 
close agreement with de Vaucouleurs et al. (1991). 
\begin{figure}
\psfig{file=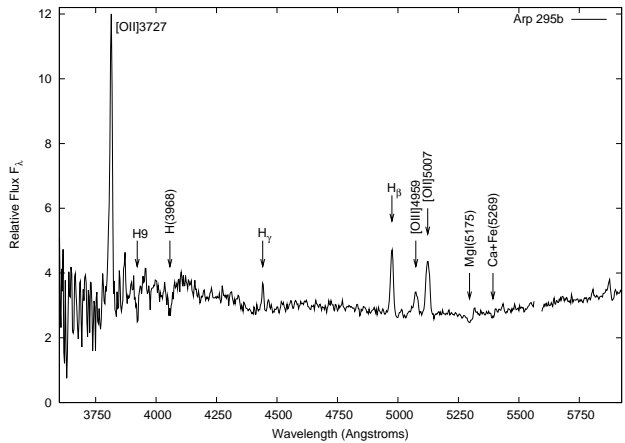,width=85mm,angle=-90}
\caption{Observed spectrum of Arp 295b, from 7200s of Bolitas data (6 exposures),
 detected lines labelled.}
\end{figure}

An error function $\sigma(\lambda)$ was estimated from the scatter between  spectra separately extracted  from each of the 1200s exposures. 
At $\lambda>4500\rm \AA$ the detection of the continuum in each pixel is 
be $\sim 20\sigma$. The error for each line's width/flux measurements was estimated by summing 
$\sigma(\lambda)$ in quadrature over the line's FWHM
of $14\rm \AA$.

Table 1 lists the detected lines with equivalent widths, measured using IRAF `splot' and corrected to restframe (by dividing by $1+z$), and for emission lines, fluxes estimated using the above calibration (not corrected for Galactic extinction).
\begin{table}
\begin{tabular}{lcc}
\hline
Line & EW $\rm \AA$& Flux ($10^{-14}$ \\
\smallskip
     & (restframe) & ergs $\rm cm^{-2}s^{-1}$) \\     
$\rm [OII]3727$ & $36.4\pm 2.1$ & $34.9\pm2.0$ \\
H9(3835.4) & $3.1\pm 1.3$ & - \\
H(3968.5) & $3.2\pm 0.9$ & - \\ 
$\rm H\gamma$ & $3.0\pm 0.4$ & $2.6\pm 0.3$ \\
$\rm H\beta$ & $10.7\pm 0.4$ & $9.2\pm 0.3$ \\
$\rm [OIII]4959$ & $4.3\pm 0.4$ & $3.7\pm 0.3$ \\
$\rm [OIII]5007$ & $11.1\pm 0.5$ & $9.7\pm 0.5$ \\
MgI(5175.4) & $1.1\pm 0.4$ & - \\
$\rm Ca+Fe(5269)$ & $0.8\pm 0.3$ & - \\
\hline
\end{tabular}
\end{table}

The OII flux corresponds to a luminosity $4.31\times 10^{41}$   ergs $\rm s^{-1}$, which from
the relation of Kennicutt (1998), $\rm SFR= 1.4\times 10^{-41}L_{\rm [OII]}$, gives the SFR as $\rm 6.03\pm 0.35~M_{\odot}yr^{-1}$. With a correction for Galactic extinction, the [OII] luminosity would be about $5.0\times 10^{41}$ ergs $\rm s^{-1}$ and thus the corresponding SFR 
$\rm 7.0\pm 0.4~M_{\odot}yr^{-1}$.

Kobulnicky et al. (2003) describe a method of estimating the metallicity
of star-forming galaxies using emission-line equivalent widths. Following these authors, we correct for stellar absorption by adding 
$2\rm \AA$ to the observed equivalent width of $\rm H\beta$.
Our measurements then give the ratios log(EW $R_{23})=0.61\pm 0.03$ and
log(EW $O_{32})=-0.37\pm 0.04$, and from equation 3 of Kobulnicky et al., 
the metallicity $\rm 12+log(O/H)=8.757\pm 0.030$. On the basis of the Allende Prieto, Lambert and Asplund (2001) measurement of the solar oxygen abundance,$\rm 12+log(O/H)=8.69\pm 0.05$, this is
$1.17\pm 0.15$ solar.

Higher luminosity is generally correlated with high metallicity, but with a large scatter. On a luminosity-metallicity plot, Arp 295b is well within the observed range of local galaxies. Compared to the 
fitted relation for local $\rm NFGS+K92$ galaxies (Kobulnicky et al. 2003), it is slightly and marginally -- $0.8\pm 0.5$ mag -- above the mean
$B$ luminosity for its (O/H). 

\section{Direct Imaging}
\label{sec:ima} 
Figure 2 shows our 6000s combined image of Arp 295b in E6690, which will include both the $\rm H\alpha$ line emission and $91\rm \AA$ of red continuum.  We therefore express the photon counts in terms of a continuum magnitude $m_{6690}$, in the AB system ($m_{AB}=48.60-2.5~{\rm log}~f_{\nu}$).
\begin{figure}
\psfig{file=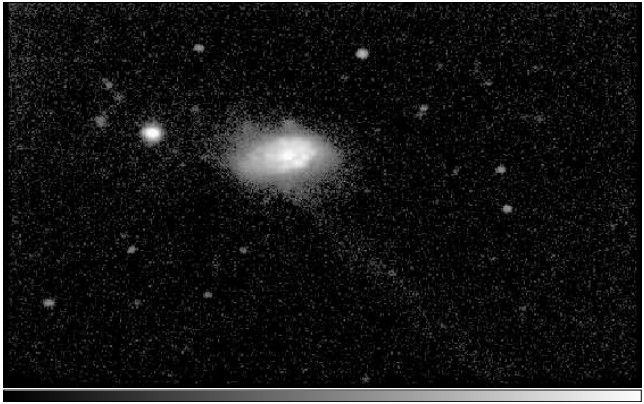,width=85mm}
\caption{Arp 295b imaged through the  E6690 filter (6000s total); the area shown is $4.71\times 2.87$ arcmin, N at the top and E at the left, with a log intensity scale.}
\end{figure}
\begin{figure}
\psfig{file=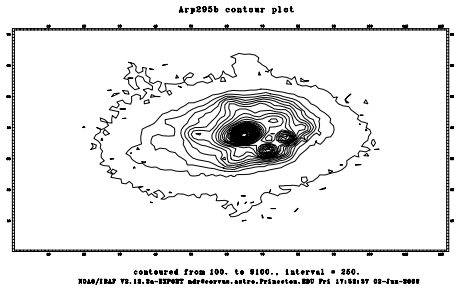,width=90mm}
\caption{Contour plot of the disk of Arp 295b, showing a $75\times 44$ arcsec area. The min and max contours correspond to intensities 22.64 and 17.84 mag
$\rm arcsec^{-2}$ in $m_{6690}$.}
\end{figure}

Figure 3 shows a contour plot of the galaxy disk. 
Arp 295b is a disturbed and somewhat asymmetric galaxy. The bright, compact nucleus is surrounded by at least 8 bright knots, presumably $\rm H\alpha$ emitting star-forming regions. The majority, and the brightest, of the knots are on the western side of the galaxy, up to 8 arcsec from the nucleus. However, the disk itself extends further to the east side of the nucleus, in the direction of the gaseous plume.
 Figure 2 also shows, faintly, the bridge extending SW towards Arp 295a, and 2 diffuse extensions north of the disk which appear to be the
`ring-shaped' $\rm H\alpha $ emission seen by Dopita et al. (2002)

To characterize the structure of the galaxy, we use IRAF `isophot.ellipse' 
to fit the E6690 image
with concentric elliptical isophotes. Figure 4 shows isophote
surface brightness (SB), ellipticity and position angle (PA) as a
function of semi-major axis ($r_{sm}$). These parameters fluctuate in the 
central $r\leq 7$ arcsec because of `knots', but 
the galaxy as a whole has an approximately exponential intensity profile with a best-fitting scalelength $r_{exp}=5.36\pm 0.55$ arcsec ($2.52\pm 0.26$ kpc). The ellipticity is consistent with the $\sim 0.43$ expected for the inclination of $55^{\circ}$ (1-cos $55^{\circ}=0.43$).

 The position angle is $\sim 100^{\circ}$ for the galaxy disk, shifting 
 to $80^{\circ}$ in the outermost isophotes where the isophotes follow the
 plume extending E of the disk. At  no radius do we find a PA as far from E-W as
 the $112^{\circ}.5$ long-slit position used by Stockton (1974). The disk itself may be
 just slightly warped, its mean PA twisting from $103^{\circ}.5$ at $r<5$
 arcsec to $98^{\circ}.3$ at 10--20 arcsec.

The total flux from Arp 295b in this passband, integrating out to 39 arcsec, is $m_{6690}=12.91$. The two UV-bright clumps reported by Neff et al. (2005)  are visible but very faint. For `clump 1', 86 arcsec east of 
the nucleus, $m_{6690}=19.41$, and for 'clump 2', 16 arcsec to its north, 
$m_{6690}=19.32$.

\begin{figure}
\psfig{file=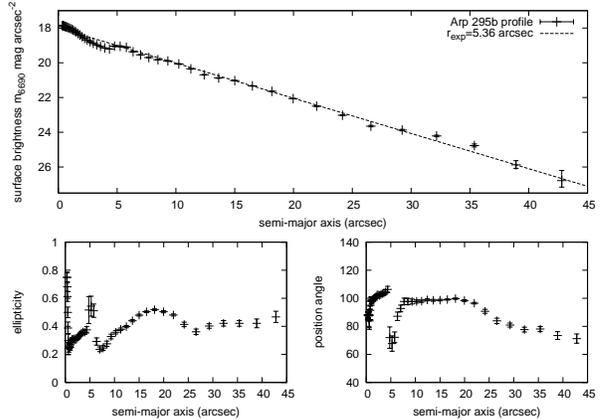,width=80mm,angle=-90}
\caption{Parameters of elliptical isophotes fit to Arp 295b, observed in E6690 : surface brightness
against semi-major axis (points) and best-fitting
 exponential (straight line); ellipticity, and position angle (in degrees
anticlockwise from the  North).}
\end{figure}

Figure 5 shows Arp 295a in E6690. Also visible is
a smaller, round, galaxy 65 arcsec NW (position angle $-59^{\circ}$) from 295a, at
RA $23^h 41^m 43^s.9$, Dec -03:39:29.  On the basis that this is probably a third galaxy in the system (which we will be able to confirm below, using MES), we refer to it hereafter as Arp 295c. 
\begin{figure}
\psfig{file=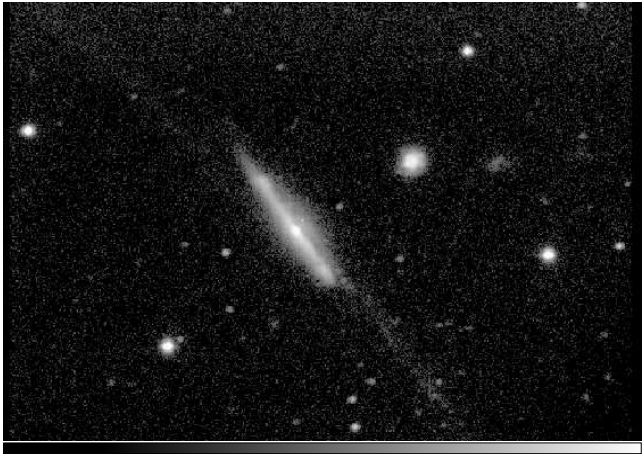,width=85mm}
\caption{Image centred on Arp 295a, E6690 filter (1200s); the area shown is $4.97\times 3.73$ arcmin, N top and E left, with a log intensity scale.}
\end{figure}
Using `isophot' we fit isophotes to both galaxies (Figure 6). Arp 295a, integrated out to 49 arcsec, has a total magnitude $m_{6690}=13.22$. The nucleus produces a central peak in the intensity plot, but at $5<r<50$ arcsec it is close to exponential with a best-fitting $r_{exp}=11.16$ arcsec, or 5.24 kpc.
The ellipticity is very high, as expected for the $\sim85^{\circ}$ inclination, and the position angle about $35^{\circ}$, consistent with the $37^{\circ}$ used by Stockton (1974).

\begin{figure}
\psfig{file=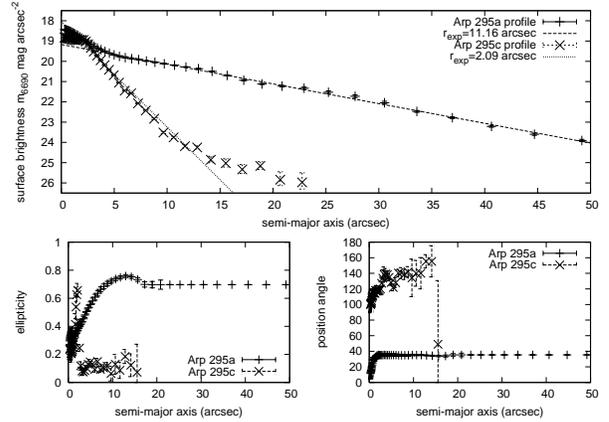,width=80mm,angle=-90}
\caption{Parameters of elliptical isophotes fit to Arp 295a and c, observed in E6690 : surface brightness
against semi-major axis (points) and best-fitting
 exponential (straight line); ellipticity, and position angle (in degrees
anticlockwise from the  North).}
\end{figure}
Arp 295c, integrated out to 23 arcsec, has a total magnitude $m_{6690}=14.65$.
The intensity profile shows an excess above an exponential at $r>12$  arcsec,
and is too flat in the centre for either an exponential or a de Vaucouleurs profile. The nucleus appears either double-peaked or barred, hence elongated,  but at all larger radii the ellipticity is very low $\sim 0.1$. The half-light radius $r_{hl}$ is 3.35 arcsec (1.57 kpc), and fitting an exponential to the steep part of the profile, $2<r<14$ arcsec, gives $r_{exp}=2.09$ arscec, or 0.98 kpc. The form of the intensity profile, the appearance and the
scalelength of Arp 295c indicate it is probably an Im type (e.g Hunter and Elmegreen 2006). 
\section{$\rm H\alpha$ flux from slitless spectra}
\label{sec:flux}
To measure the $\rm H\alpha$ line flux from each galaxy, separate from the continuum, we
take $slitless$ spectra with MES in spectroscopic mode (grating rather than mirror).
Single exposures of 1200 sec were taken on both the Arp 295b and Arp295a/c fields.
If a galaxy is a $\rm H\alpha$ emitter, these images will show the $\rm H\alpha$ line, extended in the spatial axis but also smeared out in the dispersion axis due to the
extension of the galaxy. Because no slit was used the images will contain the
entire $\rm H\alpha$ flux from all parts of the galaxy.

We see strong $\rm H\alpha$ line emission for Arp 295b and 296c,  
(confirming that 295c is equidistant, thus physically associated with the spirals). However, we see no emission line for 295a, 
so its brightness on the E6690 direct-image must have been due to its red
continuum. For 295b/c,
we extract the $\rm H\alpha$ lines from their slitless spectra, using `apall' with wide extraction apertures that encompass all the emission.

Because the galaxies are extended, wavelength calibration will inevitably be less precise when spectra are taken without slits.
Our slitless exposure of Arp 295b was taken immediately following an exposure through the slit, without moving the telescope. In this exposure the slit was 
5.77 arcsec north of the galaxy centre, so to correct for this we apply a correction $-0.98\rm \AA$ to the slitless spectrum wavelength calibration.
For our slitless spectrum of Arp 295c, the telescope was actually pointed at Arp 295a, and so we wavelength calibrate by matching the $H\rm \alpha$ line centroid to that measured on a separate exposure of Arp 295c  where the slit was used and placed on the galaxy (to look for evidence of rotation). Following these correcrion we flux calibrated using the slitless spectrum of 58 Aquilae.
Figure 7 shows integrated $\rm H\alpha$ profiles, wavelength and flux calibrated. 
\begin{figure}
\psfig{file=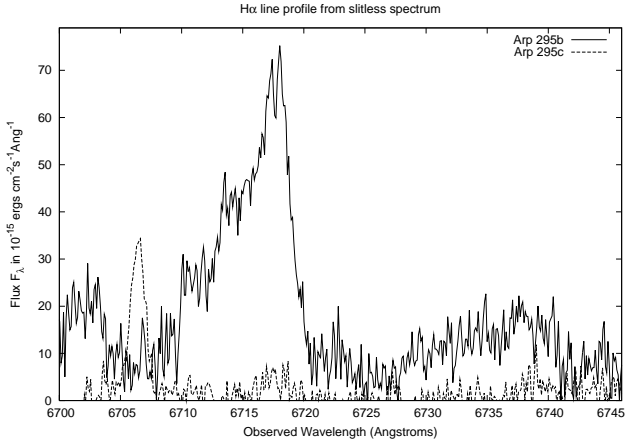,width=85mm,angle=-90}
\caption{MES spectra in the $\rm H\alpha$ line region for  the spiral Arp 295b and the irregular Arp 295c, obtained from slitless observations so as to include the entire flux from each galaxy.}
\end{figure}
We measure the total $\rm H\alpha$ flux for each galaxy by integrating
over the $\rm H\alpha$ line profile and subtracting the continuum (i.e. 
 the mean flux in wavelength intervals either side of the line), which
gives for Arp 295b, $3.47\times 10^{-13}$ ergs $\rm cm^{-2}s^{-1}$,   corresponding to a luminosity $4.27\times 10^{41}$ ergs $\rm s^{-1}$.
With a 0.102 mag correction for Galactic extinction this increases to $3.72\times 10^{-13}$ ergs $\rm cm^{-2}s^{-1}$ and $4.69\times 10^{41}$ ergs $\rm s^{-1}$. We also estimate the restframe equivalent width of $\rm H\alpha$ as $45\rm \AA$.

The centroid wavelength is $6715.61\rm \AA$, which is a redshift
0.02328 or velocity 6980 km $\rm s^{-1}$. This may be a small
 overestimate because the
receding (west) side of this galaxy is the more strongly emitting. At the nucleus or kinematic  centre 
of the galaxy the wavelength is closer to $6714.5\rm \AA$, which is a redshift
0.02312 or velocity 6930 km $\rm s^{-1}$.

We measure the $\rm H \alpha$ flux from the irregular 
Arp 295c as $5.00\times 10^{-14}$ ergs $\rm cm^{-2}s^{-1}$, which  corresponds to a luminosity $6.16\times 10^{40}$ ergs $\rm s^{-1}$.
With a 0.102 mag correction for Galactic extinction these increase to $5.49\times 10^{-13}$ ergs $\rm cm^{-2}s^{-1}$ and $6.76\times 10^{40}$ ergs $\rm s^{-1}$.
  The centroid wavelength is $6706.46\rm \AA$, which is a redshift
0.02189 or velocity 6562 km $\rm s^{-1}$.  Note the high velocity of
approach, almost 370 km $\rm s^{-1}$, relative to Arp 295b. The radial velocity relative to Arp 295a, which it is much closer to, is 287 km $\rm s^{-1}$ (295a velocity from de Vaucouleurs et al. 1991).
\section{Kinematics from $\rm H\alpha$ slit spectra}
\label{sec:kin}
Our high-resolution MES spectroscopy traces the mean line-of-sight velocity of the $\rm H\alpha$ emitting gas, as a function of distance along the slit. With multiple slit positions a 2D velocity map can be built up.

For Arp 295b we took exposures at 9 slit positions on the galaxy disk, ranging from 8.027 pixels south to 9.308 pixels north of the nucleus, always with the slit aligned E-W (The slit positions were determined using the position of a star on the corresponding `mirror-slit' images). 

 Each slit exposure was examined using IRAF `splot' to find the centroid wavelength ($\lambda_{obs}$) for the 
peak of $\rm H\alpha$ emission, as a function of position
 along the slit. Using our wavelength calibration, this position is converted into a wavelength 
$\lambda_obs$ and then to a relative velocity in the galaxy rest-frame,
 $\Delta v={c(\lambda_{obs}-\lambda_{0})}\over
{(1+z)\lambda_0}$
where $z$ is the centroid redshift of the galaxy (taken as $z=0.02312$ for
Arp 295b) and $\lambda_0=6252.8(1+z)$.

Figure 8 shows $\Delta v$ against position on the E-W axis for the 9 slit positions,
plotted as lines except for the 
slit position closest to the centre, where the velocity is plotted as points with errorbars (derived by `splot') to show typical uncertainties.
 This shows that the east side of Arp 295b 
is approaching and the west receding, by $\sim 200$ km $\rm s^{-1}$,
 and that there is a definite asymmetry in the velocity curves on the E and W sides of the galaxy, which is 
apparent at several slit positions. The greatest difference between any two velocities in this plot is 
$413.8\pm 16.3$ km $\rm s^{-1}$. Dividing this by 2 sin $55^{\circ}$ (the disk inclination), gives the maximum rotation velocity as 
$252.6\pm 9.9$ km $\rm s^{-1}$.

In addition to spectroscopy on the disk of Arp 295b, we take exposures
with the slit placed across the nucleus and disk of Arp 295a, and on one of the candidate tidal dwarf galaxies of Neff et al. (2005). However, none of these showed any visible line emission. We also observe with the slit across Arp 295c. This does show emission, but we do not find any 
significant rotation, any velocity gradient across the disk is $<10$ km $\rm s^{-1}$. This implies its orientation is almost exactly face-on.
\begin{figure}
\psfig{file=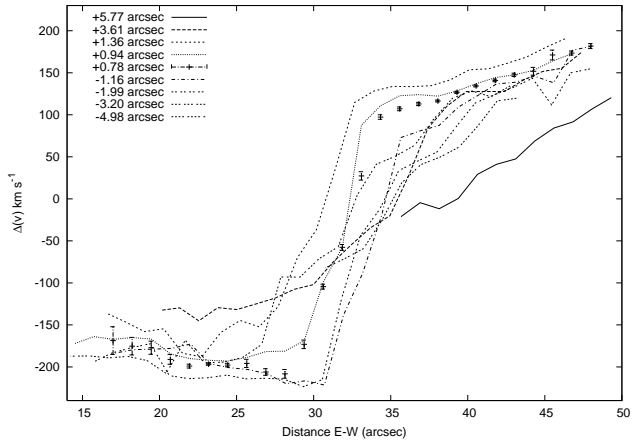,width=85mm,angle=-90}
\caption{$\rm H\alpha$ line velocity for Arp295b, in the rest-frame (relative to the approximate median recession velocity of 6930 km $\rm s^{-1}$),
 as a function of distance from E to W along each slit position.
 The slit positions are denoted by arcsec N (or S if negative) of the galaxy nucleus.
The central slit position's velocities are plotted as points with errorbars.}
\end{figure}
There are two ways in which we can try to show the kinematics more clearly.
The first is by generating a 2D velocity map (Figure 9). This shows how the 
greatest recession velocity is not at the eastern extreme but at a point
 intermediate between this and the nucleus, and that the rotation of the 
galaxy is inclined (by at least $10^{\circ}$) with respect to the
 E-W axis of our slits. 

The second way is by extracting a rotation curve along
 the true long axis of the galaxy disk. To take into account the warping
 of the disk we represent this axis as a line passing though the centre with a
PA of $103.5^{\circ}$ at $r<5$ arcsec,
 bending to a PA $98.3^{\circ}$ at larger radii (shown as the arrows on
 Figure 9). Along this locus, and 
separately in the east and west directions, we estimate the velocity at  by
interpolating between the observed values on the adjacent slit positions.
Figure 10 shows the velocity at 1 arcsec intervals, eastwards and westwards
of the nucleus. 
\begin{figure}
\psfig{file=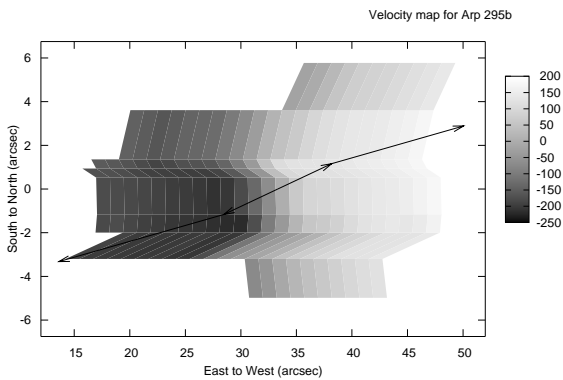,width=85mm,angle=-90}
\caption{$\rm H\alpha$ Velocity map for Arp295b in km$\rm s^{-1}$ of recession in the line-of-sight,
relative to the nucleus (6930 km $\rm s^{-1}$).  The
arrows show the approximate long axis of the disk.}
 \end{figure}
\begin{figure}
\psfig{file=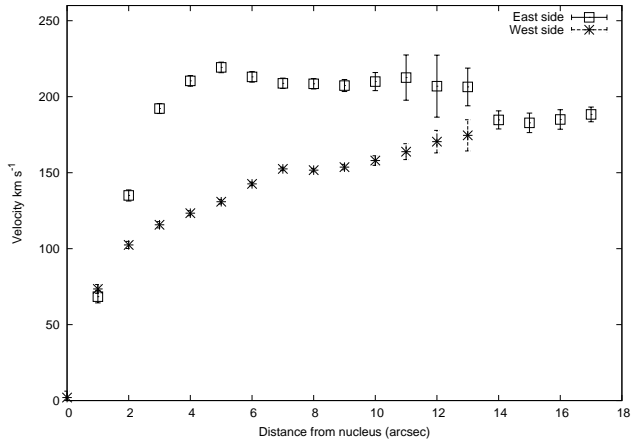,width=85mm,angle=-90}
\caption{Magnitude of the line-of-sight velocity extracted on the
 long axis of Arp295b (as depicted by the arrows on Figure 9), as a function of distance 
from the nucleus, shown separately for the east and west sides of the galaxy}.
\end{figure}
This shows clearly the asymmetry in the galaxy's rotation. On the west side the rotation curve rises slowly and monotonically on the east side it rises more steeply but begins to turn over at $r=5$ arcsec. At the smallest
($\sim 1 $ arcsec) and largest ($\sim 14$ arcsec) radii the rotaion velocities are similar on the two sides but at intermediate radii, especially around 5 arcsec, the velocity is higher on the eastern (approaching) side, the
difference reaching 88 km $\rm s^{-1}$ or a factor 1.675.

We can estimate a lower limit to the dynamical mass of Arp295b from
the rotation velocity at the largest radius observed. The mass inwards of
radius $r$ is $rv_{rot}^2/G$. At 17 arcsec, which is 7.98 kpc, 
$v_{rot}=188.3\pm 4.8/({\rm sin}~55^{\circ})=229.9\pm5.9$ km $\rm s^{-1}$, 
giving a mass within this radius $9.8\times 10^{10}M_{\odot}$.
\section{Summary and Discussion}
We have observed the Arp295 system of interacting galaxies, at $z=0.0023$, using optical spectroscopy, imaging in a narrow ($91\rm \AA$) band centred on redshifted  $\rm H\alpha$, and high-resolution spectroscopy of the 
$\rm H\alpha$ line.

Arp 295a is an almost edge-on spiral for which we measure a scalelength
$r_{exp}=5.24$ kpc in red continuum. We did not detect $\rm H\alpha$ line emission with MES, either when the slit was placed on the galaxy, or when the galaxy was observed slitless. 

Arp 295b is a spiral with a smaller scalelength $r_{exp}=2.52$ kpc.
Unlike its companion it is a strong source of $\rm H\alpha$ line emission and we measure a total $\rm H\alpha$ luminosity (corrected for Galactic extinction but not internal dust) of $4.69\times 10^{41}$ ergs $\rm s^{-1}$. Our uncorrected $\rm measurement of the H\alpha$ flux is similar to (10 per cent greater than) that given by Dopita et al. (2002). Our $\rm H\alpha$ luminosity estimate corresponds (from the relation of Kennicutt 1998) to a SFR of $\rm 3.7~M_{\odot}~\rm yr^{-1}$.

Arp 295b is asymmetric, with the disk extending further from the nucleus on the east side (away from Arp 295a), where there is also an extended plume of neutral gas (e.g. Neff et al 2005). There are bright knots of $\rm H\alpha$ emission concentrated around the nucleus
and especially on its western side. 

We obtained an optical spectrum of Arp 295b at 3600--$5900\rm \AA$, which showed emission lines of [OII]3727, [OIII]4959,5007, $\rm H\gamma$ and $\rm H\beta$, and absorption lines of H9, H(3968), MgI(5175) and $\rm Ca+Fe(5269)$.

     From the emission line flux ratios and the formula of Kobulnicky et al. (2003) we estimate a metallicity  $\rm 12+log(O/H)=8.757\pm 0.030$, approximately solar.
This is consistent with the observed metallicity-luminosity relation of spiral galaxies -- the luminosity is actually slightly ($0.8\pm0.5$ mag)
high for the (O/H). If this is significant it is presumably due to
brightening from interaction-triggered star-formation.

The galaxy's [OII] emission is very strong (equivalent width $36\rm \AA$) and if the slit spectrum is scaled to the total optical flux of the galaxy, the [OII] luminosity is 4.3 or $5.0\times 10^{41}$ ergs $\rm s^{-1}$ (without or with correction for Galactic
extinction). However, on the basis of the Kennicutt (1998) relations,  this corresponds to a SFR about twice that estimated from the $\rm H\alpha$ emission. It is possible that 
the slit sampled starbursting, presumably central, regions of the galaxy where the [OII] equivalent width was larger than for the galaxy as a whole, but it is also possible that the galaxy has a higher than average
$\rm [OII]/H\alpha$ ratio. 

The Kennicutt (1998) relation of SFR to $L_{\rm [OII]}$ does include some dust extinction. Kewley, Geller and Jansen (2004) have
more recently estimated that if $L_{\rm  [OII]}$ is corrected for (or unaffected by) dust, the $\rm SFR\times 6.58\simeq 10^{-42}L_{\rm [OII]}$. 
Hence, if Arp 295b suffers little internal dust extinction, both [OII] and $\rm H\alpha$ measures of SFR then give about 3.6 $\rm M_{\odot} yr^{-1}$. Furthermore, Kewley et al. find the intrinsic $\rm [OII]/H\alpha$ ratio to be dependent (non-monotonically) on the oxygen abundance. In the Kewley et al. relation, the $\rm [OII]/H\alpha\simeq 0.94$ we find is
 consistent with our abundance estimate, provided the internal dust extinction of Arp 295b is only $E(B-V)\simeq 0.1$--0.2 mag.

      Two 'knots' in the eastern plume were claimed by Neff et al. (2005) to be formative tidal dwarf galaxies. On our direct imaging in E6690 
we detected both objects, faintly ($R\sim 19$), but did not detect any $\rm H\alpha$ line emission with MES in spectroscopic mode, either by placing the MES slit on one knot or on a slitless observation of the whole field. Hence we could not determine whether or not these are genuine TDGs. 

    Our primary observation consists of a $\rm H\alpha$ velocity map of Arp 295b, built up by observing with MES at 9 slit positions from the north to the south edge of the galaxy disk. Arp 295b is rotating in the sense of receding on the western side, which is retrograde with respect to its orbit with Arp 295a. The maximum disk rotation velocity is
estimated as $252.6\pm 9.9$ km $\rm s^{-1}$ (consistent with the 
245 km $\rm s^{-1}$ of Stockton 1974), and from the velocity 17 arcsec
(7.98 kpc) from the nucleus we estimate the mass inward of this radius to be $9.8\times 10^{10}\rm M_{\odot}$.

Using the velocity map we determine the velocity curve on the long axis of the galaxy (taking into account that it is warped  by a few degrees), and compare the kinematics of the east and west side of the galaxy. The rotation is very asymmetric. At the smallest and largest
($r>14$ arcsec) radii, the rotation velocities are similar on both sides,
but at intermediate radii, and most strongly at $r\sim 5$ arcsec (2.35 kpc), the
rotation velocity is higher on the eastern (approaching) side by as much as
88 km $\rm s^{-1}$ in the line of sight, thus 107 km $\rm s^{-1}$ in true 
rotation velocity, or a factor 1.675.
 
We interpret this as a result of the asymmetry of the galaxy disk (Figures 2, 3), and hence of the gravitational potential, resulting from the tidal perturbation by the more massive Arp 295a. An asymmetry in the potential causes an relative asymmetry twice as great in the rotation curve
(Jog 2002).  Very close to the nucleus, the orbital kinematics will be dominated by the nucleus itself, so orbits are circular and symmetric. However, for larger orbits the effective centre of gravity is displaced eastwards of the 
nucleus, and so stars and gas orbiting the  nucleus move faster while on the eastern side (simply by Kepler's laws). Here the asymmetry in the potential would peak at $\sim 4/3$  on a circle radius 5 arcsec about the nucleus. At the largest radii the velocities on the two sides tend to converge because the offset of the centre of mass becomes small compared to the orbital radius.

On the east side the velocity curve rises rapidly to $r=5$ arcsec and beyond that is either flat or falls slightly. On the west side the velocity curve rises less steeply but monotonically to $r\simeq 13$ arcsec, and at $r\leq 7$ arcsec it approaches solid-body rotation.
We saw (Figures 2 and 3) that the brightest star-forming, $\rm H\alpha$ luminous knots tended to concentrate on the west side of the nucleus, where they extend out to 7 or 8 arcsec. 
We hypothesize that the slowed or solid-body rotation on the western side of the nucleus locally enhances the star-formation and therefore causes the
asymmetric distribution of these hotspots.

  Keel (1993), examining a sample of interacting spirals, found an association between high star-formation rates and disturbed kinematics, in particular solid-body rotation and very slow rotation (caused by near-radial stellar orbits). These altered kinematics would lower the threshold for large-scale disk instabilities to develop,
causing gas to flow inwards and triggering starbursting. Our findings for Arp 295b appear to support this model.

Of the two spirals, Arp 295a appears to be affected far less by the interaction, in that its disk is not obviously asymmetric and
it is not starbursting. While we cannot estimate the SFR of SArp 295a from our data, the far IR flux suggests its is an order of magnitude below that of the smaller spiral (Hibbard and van Gorkom 1996). The disturbed form of Arp 295b must result from the two galaxies having undergone a very close passage within the last few $10^8$ years. There are three obvious reasons why the effect on Arp 295a was different.

Firstly, although Arp 295a is only 0.1 mag brighter than 295b in the $B$-band, this magnitude will be greatly affected by internal extinction due to its edge-on orientation. In
the $K$ band (2MASS 2003) where dust extinction is less efffective, Arp 295a (with $K=10.09$) is the brighter by 0.65 mag. Similarly, Masters, Giovanalli and Haynes (2003) estimated that edge-on spirals are typically dimmed by at least $\Delta(I)\simeq 0.9$ mag relative to their face-on magnitudes. As the difference in axis ratio between Arp 295a and 295b is about 0.4 and the $B$ extinction will be about twice that in $I$, the real difference in their $B$ luminosities would be $\sim 0.1+(0.4\times2\times 0.9)=0.82$ mag. 

On this basis 295a is likely to be at least twice as massive as 295b (Stockton 1974 estimated an even higher mass ratio of 3.5, although may have underestimated the mass of 295b due to its asymmetric rotation curve). We also measure the disk scalelength of Arp 295a as larger by a factor 2.08.
The twofold or more mass difference 
means Arp 295a suffered substantially less tidal perturbation in the close passage.
Furthermore, tidal effects within the disk of Arp 295a will be less apparent due to its edge-on orientation. And despite the mass difference, Arp 295a does show definite tidal effects at larger radii in the form of the long tail extending SW.
 
Secondly, according to the HI masses of Hibbard and van Gorkon (1996), Arp 295b has a greater gas content by a factor 3.45, and this will be much more concentrated because the disk area is only about 0.23 times as great. On the basis of a $\rm SFR\propto \rho^{1.4}$ Schmidt law (Kennicutt 1998), this already could account for a factor $(3.45/0.23)^{1.4}\times 0.23=10$ in SFR. However, it is probable the greater tidal disturbance of Arp 295b is also important, 
 especially if, as in the `shock-induced star-formation' model of Barnes (2004), the local SFR is strongly dependent on the strength of tidal effects as well as on the gas density. 

Thirdly, Arp 295a is rotating prograde with respect to the mutual orbit of the galaxies, while 295b is retrograde, although with a significant tilt from the orbital plane. This could explain why tidal effects in Arp 295a produced a long thin tidal tail, while in 295b a broad gas-rich plume was ejected, much as predicted by simulations (e.g. Barnes 1992) and previously observed (e.g. Hibbard and Yun 1999) for prograde-retrograde mergers.

We also investigate a third galaxy in the system, Arp 295c, situated 65 arcsec from Arp 295a. It is rounder and much smaller than the two spirals (and approximately 1.5 mag fainter), with a half-light radius 1.57 kpc. Its radial intensity profile differs from a pure exponential, being flatter in the centre and with an excess at large radii, consistent with an Im type. We detected no significant gradient of velocity across the galaxy, implying its rotation axis is close to the line of sight (and thus perpendicular to that of Arp 295a). It is $\rm H\alpha$ luminous with 
 $L_{\rm H\alpha}= 6.76\times 10^{40}$ ergs $\rm s^{-1}$ (corrected for Galactic extinction), which corresponds to a SFR of 0.53 $\rm M_{\odot}~yr^{-1}$. 

 We measure  the recession velocity of Arp295c as
6562 km $\rm s^{-1}$, indicating a radial approach of 287 km $\rm s^{-1}$
reltive to Arp 295a. Stockton (1974) found Arp 295a to have a high disk rotation velocity of 300 km $\rm s^{-1}$. This implies the high radial velocity of Arp 295c may be accounted  by its normal orbital rotation about the massive 295a, and is not necessarily a result of the interaction of the two spirals. 

A search of the Caltech NED database revealed that a galaxy with strong ultraviolet was previously 
observed at the position of Arp 295c and named Markarian 0933
 (Markarian, Lipovetskii, Stepanian 1977; Kojoian, Elliot and Tovmassian 1981). For this, Fouque et al. (1992) measured an optical line recession velocity of 6590 km $\rm s^{-1}$, close to our measurement, and its $K_s$ band magnitude (2MASS 2003) is $K_s=13.10$, compared to $K_s=10.09$ for Arp 295a and 10.74 for Arp 295b. It is thus a factor 8.8 fainter than Arp 295b in $K_s$, compared to a factor 6.8 in $\rm H\alpha$, implying the ratios of SFR to old stellar mass are quite similar in the two galaxies. Guti{\'e}rrez et al. (2006) investigated the $\rm H\alpha$  emission from a large sample of close irregular companions of large spirals, in the same luminosity range as Arp 295c, and the $\rm H\alpha$ SFR of  0.53 $\rm M_{\odot}~yr^{-1}$ places Arp 295c above the median but near the mean level of activity for this class of galaxy. Hence it is not clear, without a detailed analysis of its spectrum, to what extent the SFR may have been boosted by the close passage of Arp 295b.   
\newpage
\acknowledgements
          NR acknowledges the support of the Universidad Nacional Autonoma de Mexico, and the help of all
	  at the Observatorio Astronomico Nacional, San Pedro Martir, in particular the very competent telescope operation of Gabriel Garcia, in the
	  observations described here.


\begin{thebibliography}
\bibitem{} Allende Prieto C., Lambert D., Asplund M., 2001, ApJ, 
556, L63.

\bibitem{} Arp H., `Atlas of Peculiar Galaxies', 1966, ApJS, 14, 1.

\bibitem{} Barnes J.E., 1992, ApJ, 393, 484.

\bibitem{} Barnes J.E., 2004, MNRAS, 350, 798.

\bibitem{} Dopita M.A., Pereira M., Kewley L.J., Capaccioli M., 2002, ApJS, 143, 47.

\bibitem{} Fouque P., Durand N., Bottinelli L., Gouguenheim L, Paturel G., 
1992, 'Catalog of Optical Radial Velocities', Observatoires de Lyon et Paris-Meudon.

\bibitem{} Giuricin G., Tektunali F.L., Monaco P., Mardirossian F, Mezzetti M., 1995, ApJ, 450, 41.

\bibitem{} Guit{\'e}rrez C.M., Alonso M.S, Funes J.G, Ribeiro M.B., 2006,
AJ, 132, 596.

\bibitem{} Hamuy M., Suntzeff N.B., Heathcote S.R., Walker A.R.,
Gigoux P., Phillips M.M.,  1994, PASP, 106, 566.

\bibitem{} Hibbard J.E., van Gorkom J.H., 1996, AJ, 111, 655.

\bibitem{} Hibbard J.E., Yun M.S, 1999, AJ, 118, 162.

\bibitem{} Hunter D.A., Elmegreen B.G., 2006, ApJS, 162, 49.

\bibitem{} Jog C.J., 2002, A\&A, 391, 471.

\bibitem{} Keel W.C., 1993, AJ, 106, 1771.

\bibitem{} Kennicutt R.C, 1998, ApJ, 498, 541.

\bibitem{} Kewley L.J., Geller M.J., Jansen R.A., 2004, AJ, 127, 2002.

\bibitem{} Kobulnicky H. A., Willmer C.N.A., Phillips A.C., Koo D.C.,
 Faber S.M., Weiner B.J., Sarajedini V.L., Simard L., Vogt N.P., 2003,
ApJ, 599, 1006.

\bibitem{} Kojoian G., Elliott R., Tovmassian H.M., 1981, AJ, 86, 811. 

\bibitem{} Markarian B.E., Lipovetskii V.A., Stepanian D.A., 1977, Astrofizika, 13, 225.

\bibitem{} Masters K., Giovanelli R. and Haynes M.P., 2003, AJ, 
126, 158.

\bibitem{} Meaburn J., 2003, Rev. Mex. A.\&A., 39, 185.

\bibitem{} Neff S.G., et al., 2005, ApJ, 619, 91.

\bibitem{} Oke J.B.,1990, AJ, 99, 162.

\bibitem{} de Souza R.E, Gadotti D.A., dos Anjos S., 2004, ApJSS, 153, 411.

\bibitem{} Stockton, A. 1974, ApJ,190, 47.

\bibitem{} de Vaucouleurs G., de Vaucouleurs A., Corwin H.G. Jr.,
 Buta R.J., Paturel G., Fouque P., 1991, 
`Third Reference Catalogue of Bright Galaxies', Springer-Verlag, Berlin Heidelberg New York.


\end{thebibliography}
\end{document}